\documentclass[aps,prd,twocolumn,showpacs,preprintnumbers,nofootinbib]{revtex4}

\usepackage{graphicx,graphics,epsfig,epstopdf,subfigure,color,psfrag}
\usepackage{hyperref}
\newcommand{\be}{\begin{equation}}
\newcommand{\ee}{\end{equation}}
\newcommand{\bea}{\begin{eqnarray}}
\newcommand{\eea}{\end{eqnarray}}
\newcommand{\nn}{\nonumber}

\newcommand{\bra}[1]{\left\langle #1 \right|}
\newcommand{\ket}[1]{\left| #1 \right\rangle}
\newcommand{\vcb}{|V_{cb}|}
\newcommand{\vub}{|V_{ub}|}

\begin{document}

\preprint{BARI-TH/709-2016}
\title{ Tension in the inclusive versus exclusive determinations of $|V_{cb}|$:\\A  possible role of new physics}
\author{P.~Colangelo and  F.~De~Fazio}
\affiliation{
Istituto Nazionale di Fisica Nucleare, Sezione di Bari, Via Orabona 4, I-70126 Bari, Italy}

\begin{abstract}
 We  reconsider the possibility that the tension in the  $\vcb$ determinations from  inclusive and exclusive $B$ decay modes is  due to a new physics effect. We
modify the Standard Model  effective Hamiltonian for semileptonic $b \to c$ transitions including a  tensor operator with a lepton flavour dependent coupling $\epsilon_T^\ell$, and
 investigate separately  the muon and electron modes.
The interference term between SM and NP,  proportional to the lepton mass,  has different impact  in the inclusive and exclusive $B$ modes to muon.
 Moreover, even when the lepton mass is small as for the electron, 
the  NP effect is different in inclusive and exclusive $B$ channels.
For both $\mu$ and $e$ we find a region of $\epsilon_T^{\mu,\,e}$ where the constraints from  $B^- \to D^{(*)0} \ell^- {\bar \nu}_\ell$ and  $B \to X_c \, \ell \, \bar  \nu_\ell$   are satisfied for the same  $\vcb$.  
 \end{abstract}
\maketitle

{\it Introduction.}
The precise determination of the Standard Model (SM) parameters is a fundamental  step towards the search for new physics (NP). In particular, the CKM matrix elements play a peculiar role, being related to the SM  description of CP violation in the quark sector.
Special cases are $V_{ub}$ and $V_{cb}$, whose ratio  $|V_{ub}|/|V_{cb}|$  appears in the length of one side of the unitarity triangle.
They can be measured in several ways,  in particular using  semileptonic decays of $b$-flavoured hadrons, and 
for both $\vub$ and $\vcb$ the results  obtained from inclusive and exclusive $B$ semileptonic modes are only marginally compatible.
From exclusive  decays,   the average   in \cite{DeTar:2015orc}
\be
\vcb_{excl} =(39.78\pm0.42)\times 10^{-3}  \label{excl}
\ee
is based on experimental data  and on hadronic quantities computed by lattice QCD
 \cite{Glattauer:2015teq,Lattice:2015rga,Aubert:2008yv,Bailey:2014tva}. 
 The  Particle Data Group  is more conservative: 
 $\vcb_{excl} =(39.2\pm0.7)\times 10^{-3}$ 
 \cite{Agashe:2014kda},  while  in a recent study of the single  $B \to D$ mode  $\vcb=(40.49 \pm 0.97) \times 10^{-3}$ is obtained \cite{Bigi:2016mdz}.

Results  from   inclusive 
$B$ 
decays are summarized  in 
\cite{Alberti:2014yda,Gambino:2016fdy,Agashe:2014kda}:
\be
\vcb_{incl} =(42.21\pm0.78)\times 10^{-3} . \label{incl}
\ee
The tension between (\ref{excl}) and (\ref{incl}), one of  the so-called  {\it flavour anomalies} \cite{Ricciardi:2016pmh},   affects  the SM predictions for several other  observables, as discussed, e.g., in \cite{Blanke:2016bhf}.

Another flavour anomaly are  the ratios 
${\cal R}(D^{(*)})=\frac{{\cal B}(B \to D^{(*)} \tau {\bar \nu}_\tau)}{{\cal B}(B \to D^{(*)} \mu {\bar \nu}_\mu)}$, that exceed the SM prediction   \cite{Lees:2012xj,Aaij:2015yra,Abdesselam:2016cgx}.\footnote{A recent analysis with  references to previous  studies is   in
\cite{Becirevic:2016hea}.}  Surprisingly, both the $\vcb$ and ${\cal R}(D^{(*)})$ anomalies  involve tree-level processes, and the second one  points to violation of lepton flavour universality (LFU).
In \cite{Biancofiore:2013ki} it has been shown
  that  an additional tensor operator in the effective Hamiltonian inducing semileptonic $b \to c$  decays could accommodate the experimental findings for ${\cal R}(D^{(*)})$,
 assuming that it contributes only for  $\tau$ leptons. The coupling  $\epsilon_T^{\tau}$ of  the new term in the effective Hamiltonian is constrained by the  $R(D^{(*)})$ data  in a region  allowing to reconcile the theoretical result with measurement. 
Hence, it is worth scrutinizing if this kind of violation of LFU
  is at work also in the case of   the  $\vcb$ anomaly.
 \\ \indent
The possibility that new physics  is involved in the difference between inclusive and exclusive determinations of $\vcb$ has been previously considered  \cite{Faller:2011nj,Crivellin:2014zpa,Bordone:2016tex}. It has been excluded using the argument that
a new scalar or a new tensor  operator in the effective $b \to c$ Hamiltonian, for   massless leptons, produces the same effect in both exclusive and inclusive semileptonic modes at zero recoil, inducing  the same changes  in $\vcb$.  Moreover,  introducing a new vector or axial-vector structure in the Hamiltonian leads to  modified $W$ couplings, and this,   due to  the  $SU(2)$  symmetry of SM,  in turn modifies  the $Z$ couplings to fermions at a level experimentally excluded \cite{Crivellin:2014zpa}. \\
\indent
However,  in the inclusive $B$ mode the determination of $\vcb$ 
goes through 
the measurement of the moments of the full 
lepton energy spectrum and of the hadronic mass distribution,  to obtain  the OPE parameters   in the theoretical expression of the inclusive width. $\vcb$ is then determined  comparing the  theoretical and experimental  width   \cite{Gambino:2015ima}. 
This differs from the exclusive modes that are analyzed close to  zero recoil  \cite{Agashe:2014kda}. As we show for the new tensor operator,  
the lepton mass, in the case of muons,  
can be
 large enough to produce a sizable interference  between the SM and the NP contribution, with  different effects in the exclusive  and in the inclusive $B$ decays  and a different  impact on  $\vcb$,  jeopardizing the argument in \cite{Crivellin:2014zpa}.
 For this reason, whenever possible we consider separately the  muon and electron modes. 
Moreover,  we find that the NP effect is not the same in the inclusive and exclusive channels, and this also in  the  electron case where the interference between SM and NP is negligible.
\\ \indent
To be specific, we extend the  effective  Hamiltonian governing the  $b \to c \ell \bar \nu_\ell$ transitions 
as follows
\cite{Becirevic:2012jf,Tanaka:2012nw,Biancofiore:2013ki}:
\bea
H_{eff}=  {G_F \over \sqrt{2}}V_{cb}&& \hskip-0.4cm\Big[ {\bar c} \gamma_\mu (1-\gamma_5) b \, {\bar \ell} \gamma^\mu (1-\gamma_5) {\bar \nu}_\ell \nn \\&& \hskip-0.4cm+ \epsilon_T^\ell \, {\bar c} \sigma_{\mu \nu} (1-\gamma_5) b \, {\bar \ell} \sigma^{\mu \nu} (1-\gamma_5) {\bar \nu}_\ell \Big] \, . \label{heff}
\eea
We assume that  not only $\epsilon_T^\tau$ \cite{Biancofiore:2013ki}, but also $\epsilon_T^{(\mu,e)}$ can be different from zero,
and we  investigate the effect of  the new term  for light leptons and  the consequences for  $\vcb$.

 {\it Inclusive $B\to X_c \ell \bar \nu_\ell$ decays.}
The heavy quark expansion (HQE) \cite{Chay:1990da,Bigi:1992su,Blok:1993va,Manohar:1993qn} allows to write the  inclusive decay width
of  heavy hadrons ($H_Q$)   as a series in powers of  the inverse heavy quark mass. Each term is the product of coefficient functions times the   $\bra{H_Q} O_i \ket{H_Q} $ matrix elements  of local operators $O_i$ of increasing dimension; at each order in $1/m_Q$, the coefficient functions 
can be further  expanded in $\alpha_s$. The leading term  corresponds to the  free $Q$ decay width; the ${\cal O}(m_Q^{-1})$  term is absent. 
Specializing to  $B$ meson,  the  
expression of  
$B \to X_c \ell \bar \nu_\ell$ decay width  in SM for massless leptons, with  ${\cal O}(\alpha_s^2)$ and  ${\cal O}(1/m_b^3)$ corrections,  can be found in \cite{Alberti:2014yda}.
Here we compute $\Gamma(B \to X_c \ell \bar \nu_\ell)$ including the contribution of the tensor operator in  Eq.~(\ref{heff}) and considering massive leptons. The decay distribution in the dilepton invariant mass  ${\hat q}^2=q^2/m_b^2$ reads
\be
\frac{d\Gamma}{d\hat q^2}=C(q^2) \left[
\frac{d\tilde \Gamma}{d\hat q^2}|_{SM}+ |\epsilon_T|^2 \frac{d\tilde \Gamma}{d \hat q^2}|_{NP}+{\rm Re}(\epsilon_T)
\frac{d\tilde \Gamma}{d\hat q^2}|_{INT}\right]  ,  \label{incl-dgamma} 
  \ee
  with
  $C(q^2)= \frac{G_F^2 |V_{cb}|^2 m_b^5}{96 \pi^3} \lambda^{1/2}\left(1-\frac{{\hat m}_\ell^2}{{\hat q}^2}\right)^2$,
   ${\hat m}_\ell=m_\ell/m_b$,  $\lambda=\lambda(1,\rho,{\hat q}^2)$ the triangular function and $\rho=m_c^2/m_b^2$.   Using the HQE,
each term  A=SM, NP, INT  in (\ref{incl-dgamma})   can be  written   as
  \be \hspace{-0.2cm}
  \frac{d\tilde \Gamma}{d\hat q^2}|_A=\frac{d\tilde \Gamma_0}{d\hat q^2}|_A-\frac{\mu_\pi^2-\mu_G^2}{2 m_b^2}\frac{d\tilde \Gamma_{1/m_b^2}^{ (1)}}{d \hat q^2}|_A+\frac{\mu_G^2}{ m_b^2}\frac{d\tilde \Gamma_{1/m_b^2}^{ (2)}}{d \hat q^2}|_A \, . \label{incl-dgtilde}
  \ee
The leading order terms  in the $1/m_b$ expansion (\ref{incl-dgtilde}) read:
    \bea
\frac{d\tilde\Gamma_0}{d\hat q^2}|_{SM}=&&
(1-\rho)^2 \left(1+2\frac{{\hat m}_\ell^2}{{\hat q}^2}\right)+\nn \\&&
 (1+\rho) {\hat q}^2 \left(1-\frac{{\hat m}_\ell^2}{{\hat q}^2} \right) - {\hat q}^4 \left(2+\frac{{\hat m}_\ell^2}{{\hat q}^2}\right)\,\, , 
\nn \\
\frac{d\tilde \Gamma_0}{d\hat q^2}|_{NP}=&& 8 \left(1+2 \frac{{\hat m}_\ell^2}{{\hat q}^2} \right)\left[2(1-\rho)^2-{\hat q}^2(1+\rho+{\hat q}^2) \right]  \,\, , \nn
  \\
\frac{d\tilde \Gamma_0}{d\hat q^2}|_{INT} =&&
-36 \, \sqrt{\rho} \,{\hat m}_\ell (1-\rho+{\hat q}^2) \,\,\, . \label{dgdq20}
\eea
For the $1/m_b^2$ corrections we find:
\be
\frac{d\tilde \Gamma_{1/m_b^2}^{ (1)}}{d\hat q^2}|_A=\frac{d\tilde \Gamma_0}{d\hat q^2}|_A \hskip 0.7cm (A=SM, \, NP, \, INT), 
\ee
\begin{widetext}
\bea
\frac{d\tilde \Gamma_{1/m_b^2}^{ (2)}}{d\hat q^2}|_{SM}&=&2\Bigg\{\hskip -0.2cm-\lambda\left(2+\frac{{\hat m}_\ell^2}{{\hat q}^2}\right)-[-3+8\rho-3\rho^2+3{\hat q}^2(1+\rho)]\left(1+\frac{{\hat m}_\ell^2}{{\hat q}^2}\right) -\frac{{\hat m}_\ell^2}{{\hat q}^2}(3-\rho)\nn \\
&+&\frac{1}{\lambda}
\left[(1-\rho)^2 \left[3\rho\left(1+ \frac{{\hat m}_\ell^2}{{\hat q}^2} \right)-\left(1- \frac{{\hat m}_\ell^2}{{\hat q}^2} \right) \right]
+{\hat q}^2 \left[2(1+3\rho)-(1+12\rho+3\rho^2)\left(1+\frac{{\hat m}_\ell^2}{{\hat q}^2}\right)\right]\right]\Bigg\}  , \nn \\
\frac{d\tilde \Gamma_{1/m_b^2}^{ (2)}}{d\hat q^2}|_{NP}&=&-16\left(1+2\frac{{\hat m}_\ell^2}{{\hat q}^2}\right)
\left\{\lambda+\rho(5-3\rho)+3{\hat q}^2(1+\rho) 
+\frac{1}{\lambda} \left[{\hat q}^2(-1+6\rho+3\rho^2)-(1-\rho)^2(-1+3\rho) \right] \right\}   , \,\,\,\,\,\label{powerNP}  \\
\frac{d\tilde \Gamma_{1/m_b^2}^{ (2)}}{d\hat q^2}|_{INT}&=&24\sqrt{\rho}\, {\hat m}_\ell \left\{-2-3{\hat q}^2+3\rho+\frac{1}{\lambda} \left[3(1-\rho)^2-{\hat q}^2(1+3\rho)\right]\right\}  . \nn
\eea
\end{widetext}
$\mu_\pi^2$ and $\mu_G^2$ parameterize the matrix elements $\bra{ B} O_i \ket{B}$  of dimension 5 operators.
Setting $\epsilon_T=0$ we recover the SM result for massive leptons  \cite{Falk:1994gw}.
In our numerical analysis we also include  $1/m_b^3$  and  ${\cal O}(\alpha_s^2)$ corrections in  the SM contribution,  neglecting  the lepton mass  \cite{Alberti:2014yda}.
We do not include such corrections in the NP term, which has  to be small with respect to the  SM one. Hence, we write the inclusive semileptonic width as follows: 
\bea
\Gamma_{sl}&=&\left (\int_{\hat q^2_{min}}^{\hat q^2_{max}}\,\frac{d\Gamma}{d\hat q^2} d\hat q^2 \right)  +\Gamma_0 \Big[a^{(1)}\frac{\alpha_s(m_b)}{\pi} \nn \\ 
&+&a^{(2,\beta_0)}\beta_0\left(\frac{\alpha_s}{\pi} \right)^2 +a^{(2)}\left(\frac{\alpha_s}{\pi} \right)^2 +p^{(1)}\frac{\alpha_s}{\pi} \frac{\mu_\pi^2}{m_b^2}  \nn \\
&+& g^{(1)}\frac{\alpha_s}{\pi}  \frac{\mu_G^2}{m_b^2}  + d^{(0)} \frac{\rho_D^3}{m_b^3}-g^{(0)} \frac{\rho_{LS}^3}{m_b^3} \Big]\,\,. \label{opeW}
\eea
The first integrand is in Eq.(\ref{incl-dgamma}).   $\Gamma_0$ is the LO SM width, $\rho_D^3$, $\rho_{LS}^3$ parametrize dimension 6 operator matrix elements, and the various coefficients are reported in \cite{Alberti:2014yda}. We use 
 their  values  in the kinetic scheme, while for $m_c$  we use the $\overline{MS}$  result at the scale $\mu=3$ GeV, as  in  \cite{ Gambino:2013rza}.
\begin{figure}[t!]
\vspace*{-0.2cm}
\includegraphics[width = 0.237\textwidth]{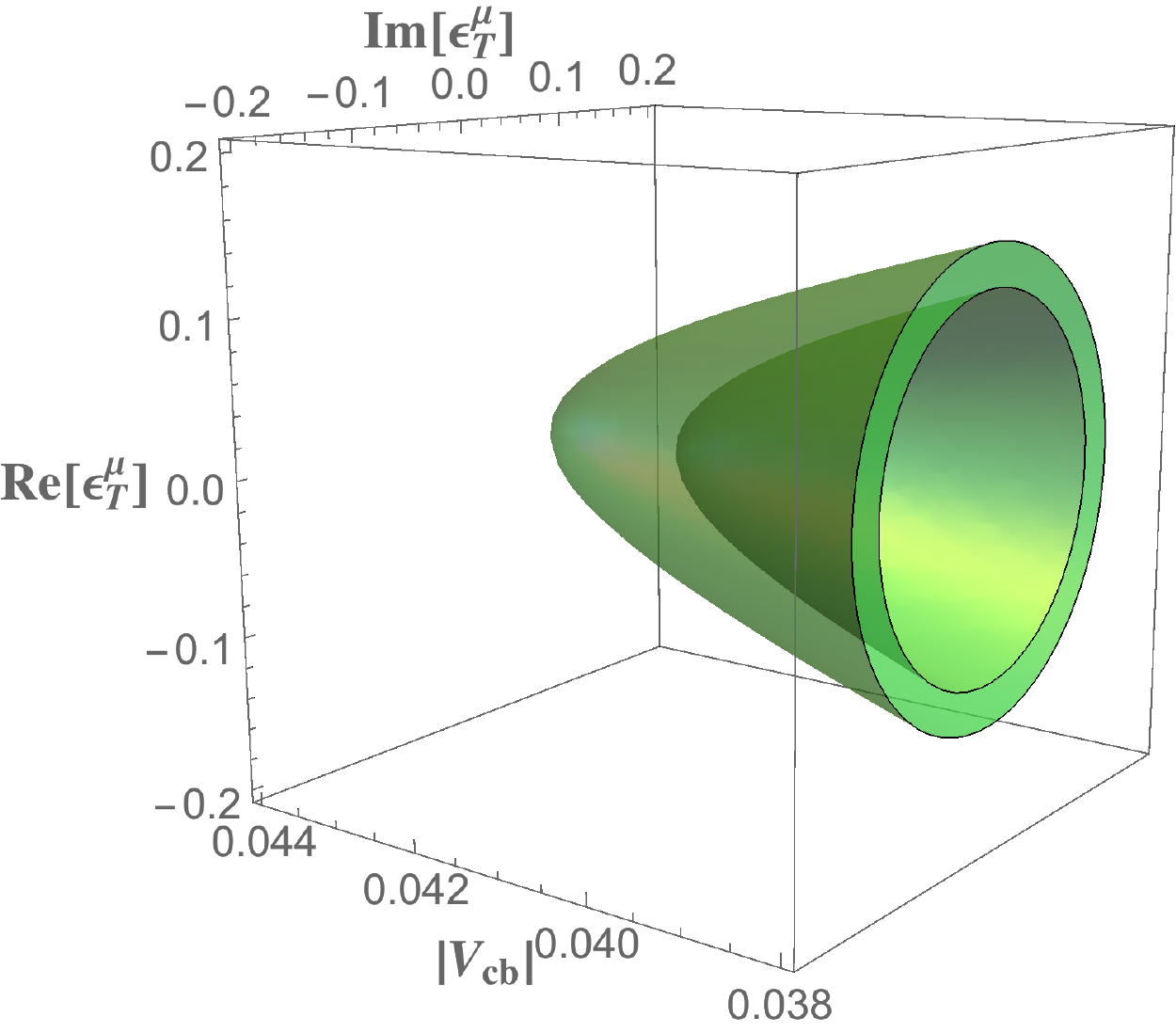}
 \includegraphics[width = 0.237\textwidth]{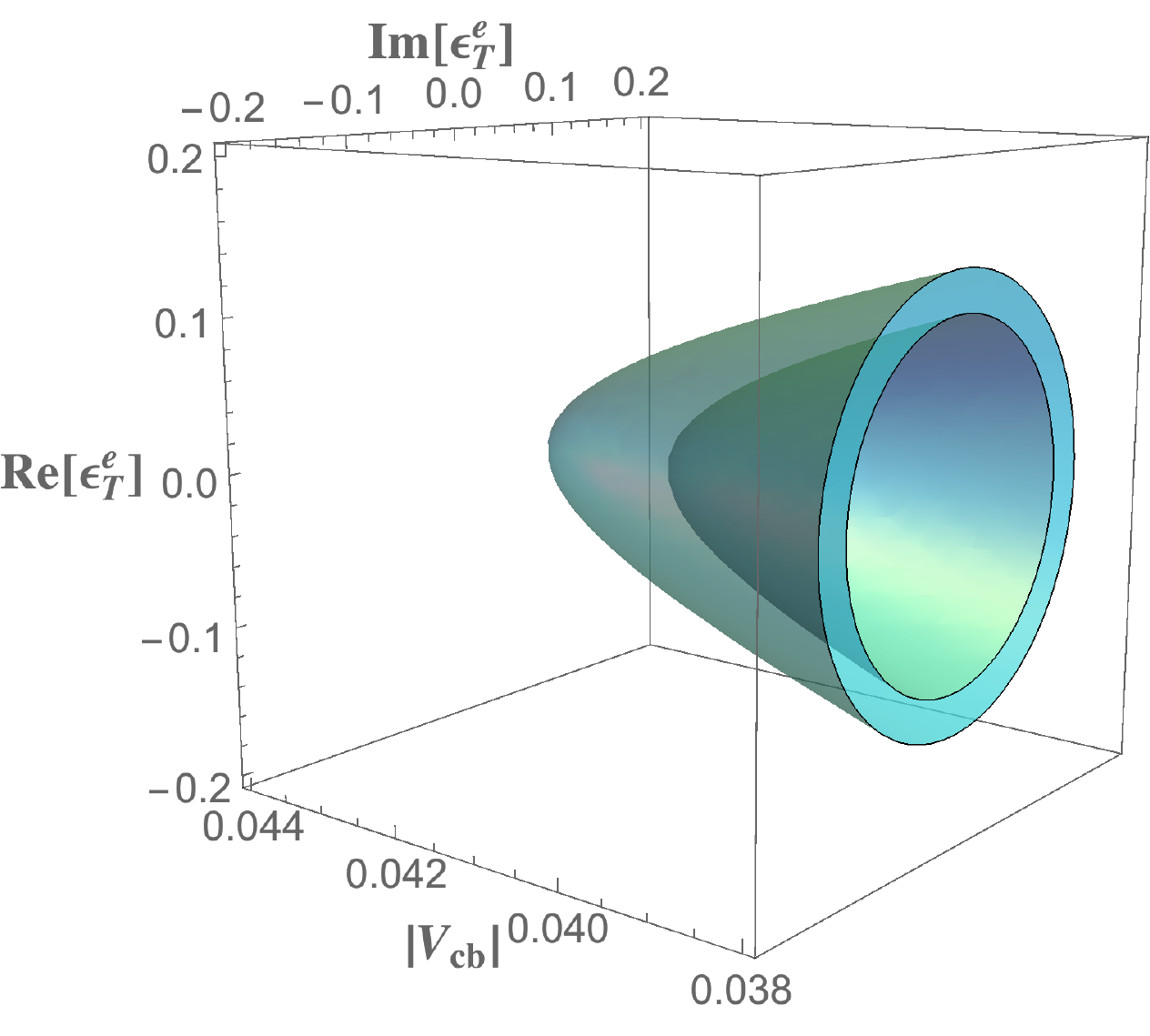}
\caption{Cutaway view of the parameter space ${\rm Re}(\epsilon_T^\ell)$, ${\rm Im}(\epsilon_T^\ell)$, and $\vcb$ allowing to obtain  ${\cal B}(B^+ \to X_c \ell^+ \nu_\ell)$ within $1 \sigma$ for muon (left) and electron mode (right).}\label{fig:incl}
\end{figure}
The Particle Data Group  quotes \cite{Agashe:2014kda}:
\be
{\cal B}(B^+ \to X_c e^+ \nu_e) = (10.8 \pm 0.4) \times 10^{-2} . \label{charged} \\
\ee
Since no data are separately reported for muon, we use (\ref{charged}) also in that case.
\begin{figure}[h!]
\vspace*{-0.2cm}
\includegraphics[width = 0.34\textwidth]{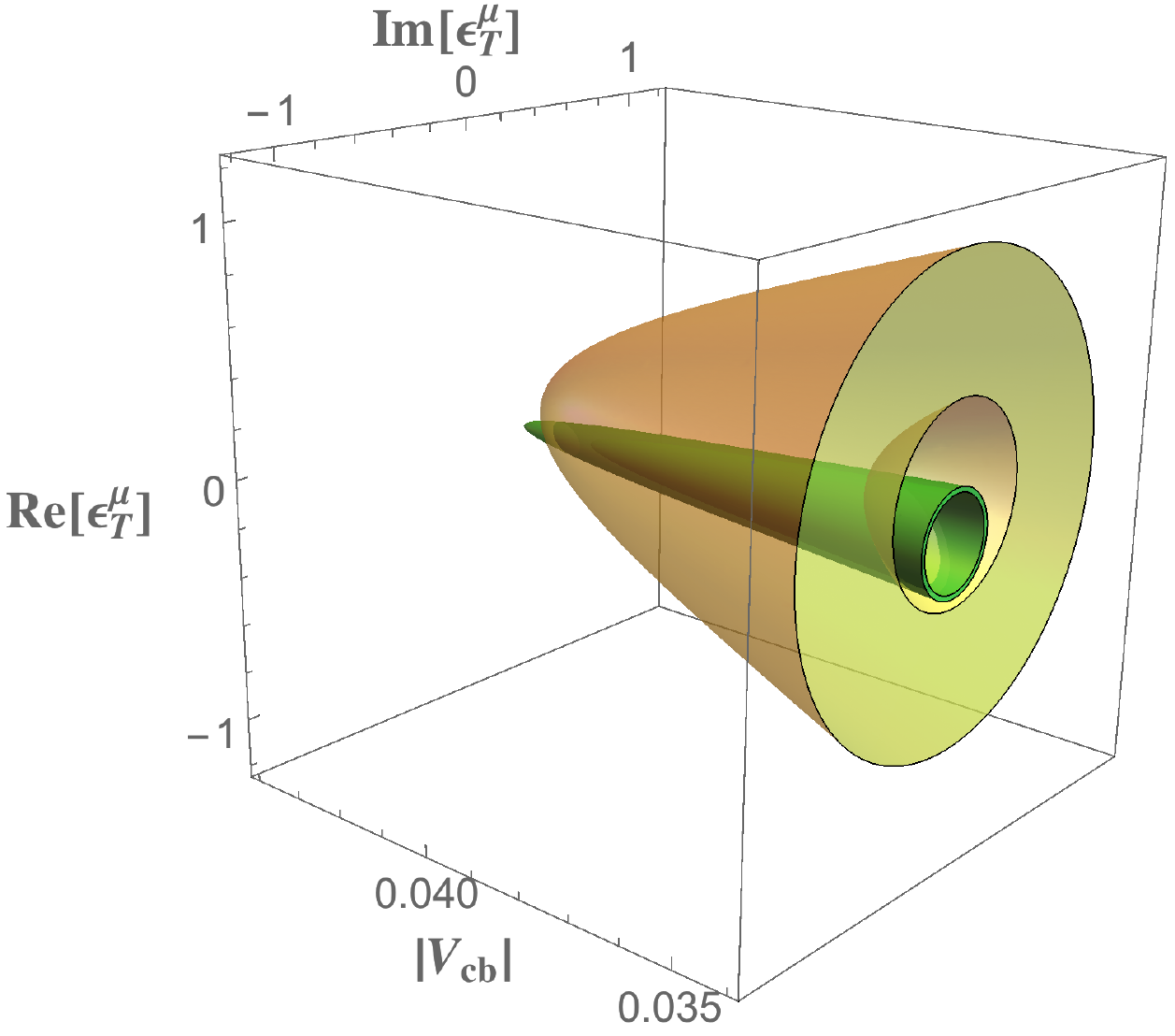}\\
\includegraphics[width = 0.34\textwidth]{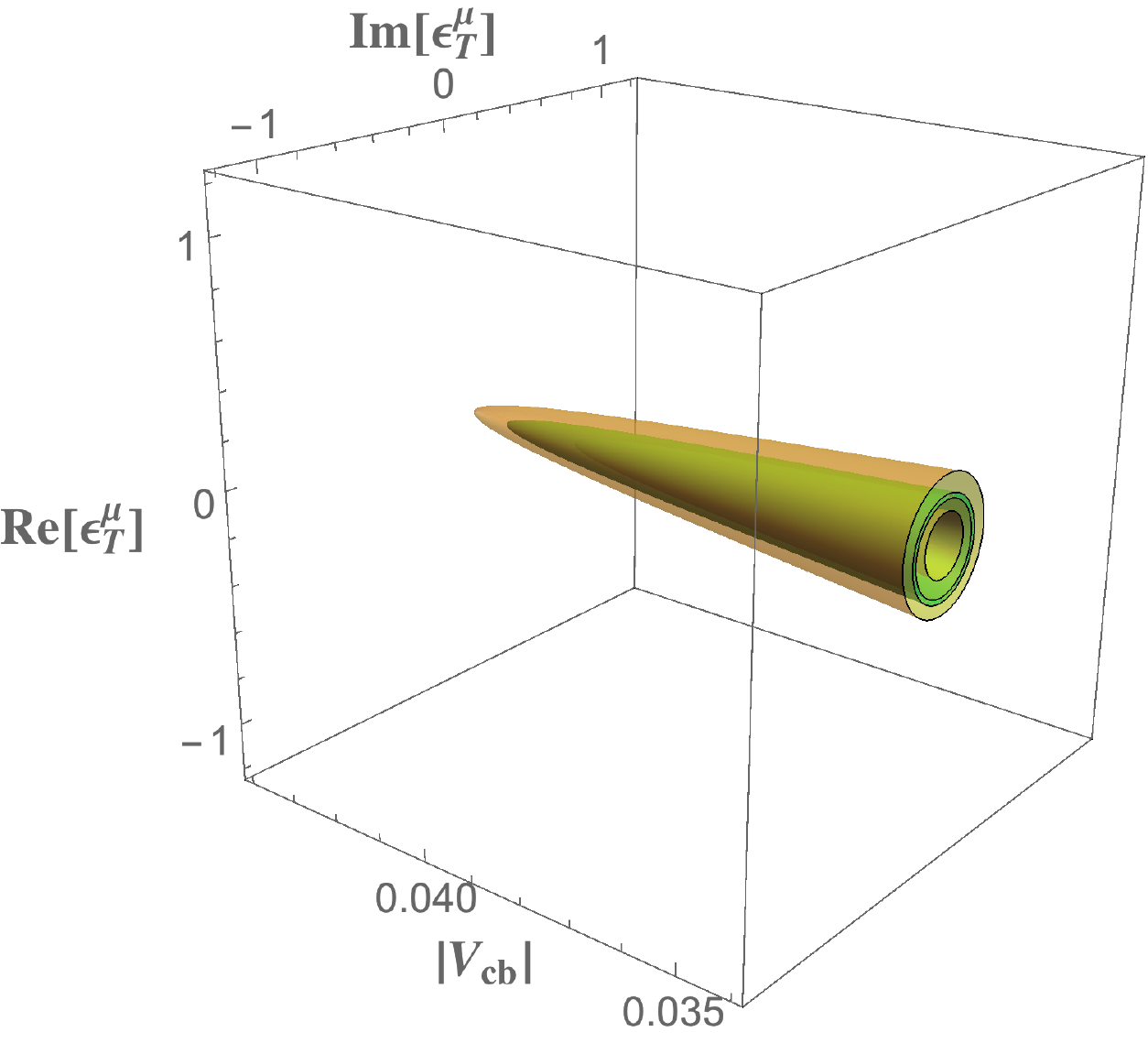}\\
\includegraphics[width = 0.34\textwidth]{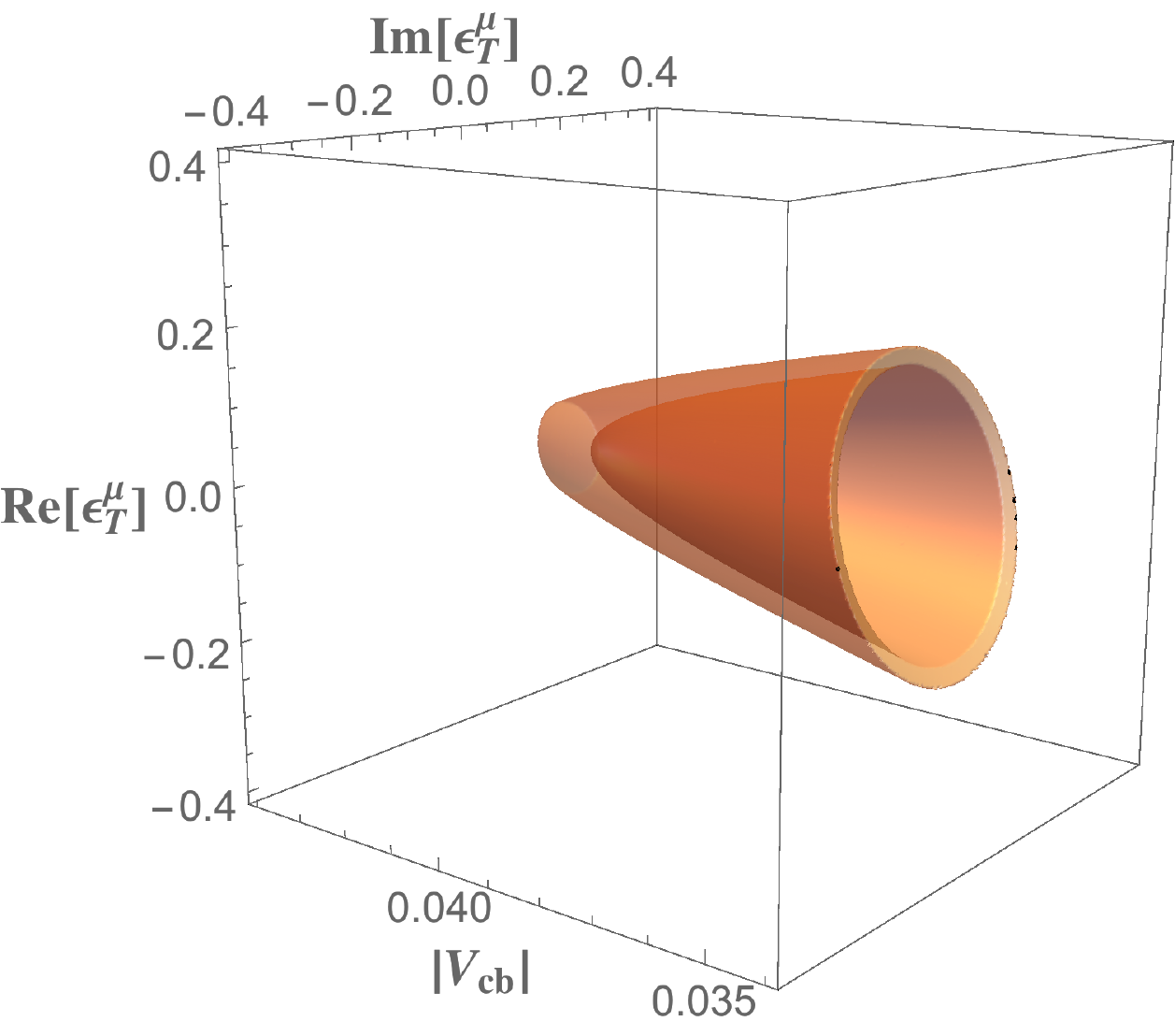}
\caption{Muon modes:  allowed regions in the $({\rm Re}(\epsilon_T^\mu),\,{\rm Im}(\epsilon_T^\mu),\,\vcb)$ parameter space, determined from  the inclusive mode (green hollow region) together with  the  exclusive  $D$ mode (yellow hollow region, upper plot), and from  the inclusive mode together with the decay to $D^*$   (yellow hollow region, middle plot).   
The intersection  of the three regions is shown in the lower plot. }\label{fig:comb-mu}
\end{figure}
In Fig.~\ref{fig:incl} we show the result of exploiting the datum Eq.~(\ref{charged}) at  1$\sigma$  level to  constrain $({\rm Re}(\epsilon_T^\ell),\,{\rm Im}(\epsilon_T^\ell),\,\vcb)$ for  the  muon and  electron mode. We keep the lepton masses distinct in the two cases.
The inclusive branching ratio bounds $\vcb$  from above: 
imposing the 1$\sigma$ constraint,  one finds   $\vcb \le 42.85 \times10^{-3}$ in the muon case,  and $\vcb \le 42.73 \times10^{-3}$
in the electron case, in correspondence to the SM point  $\epsilon_T^\ell=0$. When $\epsilon_T^\ell$ is allowed to deviate from zero, the hollow regions in Fig.~\ref{fig:incl} are found, which  continue  to smaller values of $\vcb$ if $|\epsilon_T^\ell |$ is increased; however,   a lower bound on $\vcb$ is set by the exclusive modes.

{\it Exclusive $B \to D^{(*)}$ modes.}
\begin{figure}
\includegraphics[width = 0.49\textwidth]{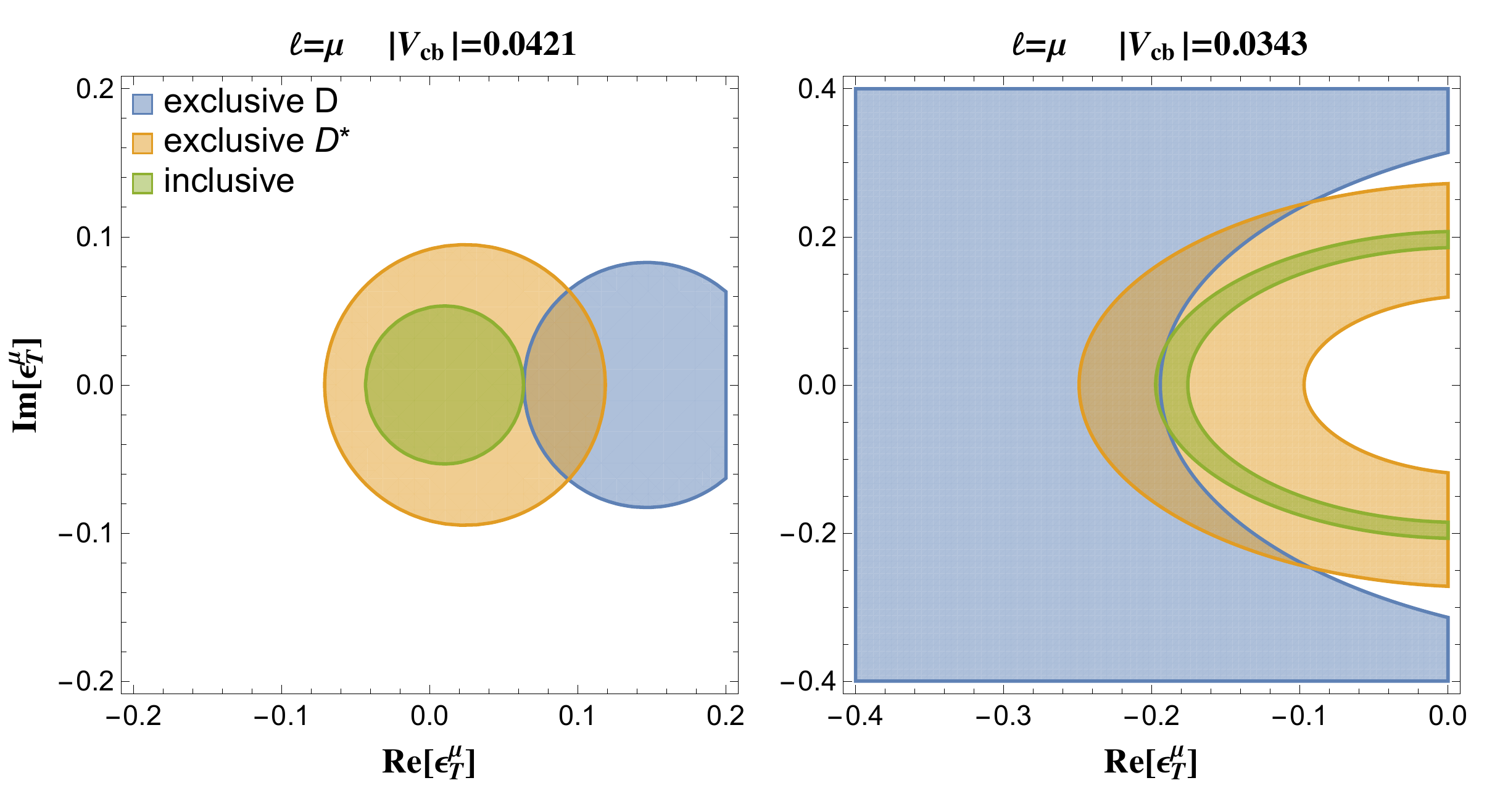}\\
\caption{Muon channel: projections of the overlap parameter region in Fig.~\ref{fig:comb-mu} on the $({\rm Re}(\epsilon_T^\mu),\,{\rm Im}(\epsilon_T^\mu))$ plane. In each panel, the blue region corresponds to the constraint from the exclusive decay to $D$, the orange region  to the constraint from  $D^*$, and the green region to the constraint from the inclusive mode. The left panel corresponds to the largest allowed  value of $\vcb$,  the right panel  to the smallest allowed value of $\vcb$. Outside this range of $\vcb$, the  parameter regions bound from the three decay modes do not overlap. }\label{rpnew-mu}
\end{figure}
The theoretical description  of $B \to D^{(*)} \ell \bar \nu_\ell$ modes requires the  $B\to D^{(*)}$ form factors. In the HQ limit all such form factors  can be expressed in terms of  
the Isgur-Wise function $\xi(w)$ \cite{Isgur:1989ed}, with $w$  related to the dilepton invariant mass,  $q^2=m_B^2+m_{D^{(*)}}^2-2 m_B m_{D^{(*)}} w$. Corrections involve both $1/m_Q$  and ${\cal O}(\alpha_s)$ terms,  and can be found, e.g.,  in \cite{Neubert:1993mb,Caprini:1997mu}.  For finite quark mass, several form factor determinations are available. For  $F_1(q^2)$ and $F_0(q^2)$  in the $B \to D$ matrix element of the weak current 
 $j_{\mu}=\bar c \gamma_{\mu}(1-\gamma_5) b$, the lattice QCD results in
   \cite{Lattice:2015rga}  have been used to obtain the average Eq.~(\ref{excl}). For consistency, we use them.
Moreover, we determine the form factor $F_T(q^2)$  parametrizing the matrix element of the tensor current, $j_{\mu \nu}=\bar c \sigma_{\mu \nu}(1-\gamma_5) b$,
 from $F_1(q^2)$  and the HQ relation at  NLO  \cite{Neubert:1993mb,Caprini:1997mu}, following the procedure described in details in \cite{Biancofiore:2013ki}.
The standard parameterization of the   $B \to D^*$   matrix element  of the currents $j_\mu$ and $j_{\mu \nu}$  involves the form factors $V$, $A_i$ ($i=0,3)$,  and $T_i$ ($i=0,5)$, as defined in 
\cite{Biancofiore:2013ki}. Lattice QCD results  are only available  for $A_1$ at the zero recoil $w=1$ \cite{Bailey:2014tva}: we determine the other ones using again HQ relations at  NLO  \cite{Biancofiore:2013ki}. 

Starting from $H_{eff}$ in (\ref{heff}),
 the differential  $B \to M_c \ell {\bar \nu}$ decay distribution, with  $M_c=D,D^*$  can be written as
\be
{d \Gamma \over dq^2}(B \to M_c \ell \bar \nu_\ell)= {d  \Gamma \over dq^2}\Big|_{SM}+{d  \Gamma \over dq^2}\Big|_{NP}+{d  \Gamma \over dq^2}\Big|_{INT} , \label{dgammadq2-generic}
\ee
with the three terms for massive leptons given in \cite{Biancofiore:2013ki}. 

For  the modes  $B^- \to D^0 \ell^- \bar \nu_\ell$,
 the BaBar Collaboration has reported  separate measurements  for $\mu$ and $e$ \cite{Aubert:2008yv}:
\bea
{\cal B}(B^- \to D^0 \mu^- \bar \nu_\mu) &=& (2.25 \pm 0.04 \pm 0.17) \times 10^{-2} \,\,\,\,\,  \label{babar-mu} \\
{\cal B}(B^- \to D^0 e^- \bar \nu_e) &=& (2.38 \pm 0.04 \pm 0.15) \times 10^{-2}  . \,\,\,\,\,\,\,\,   \label{babar-el}
\eea
We constrain $({\rm Re}(\epsilon_T^\ell),\,{\rm Im}(\epsilon_T^\ell),\,\vcb)$
 comparing  these data with the theoretical expressions 
 including the errors on the  form factors quoted in \cite{Lattice:2015rga}, and the uncertainty on the parameter $\bar \Lambda=m_{H_Q}-m_Q$  entering in the HQ relation between $F_1$ and $F_T$. We  use the  conservative  value $\bar \Lambda=0.5 \pm 0.2 $ GeV.

In the case of  $B^- \to D^{*0} \ell^- \bar \nu_\ell$, we consider the distribution $\displaystyle{\frac{d \Gamma}{dw}}$ and compare the theory prediction to experiment close to the zero recoil point $w \to 1$. 
The BaBar Collaboration fits the measured data using the expression 
\bea
&&{d \Gamma \over dw}(B \to D^* \ell \bar \nu_\ell)= {G_F^2 |V_{cb}|^2  m_B^5 \over 48 \pi^3} (1-r^*)^2 {r^*}^3 W_{D^*}(w)  \nn \\
 &&h_{A_1}^2(w) \sqrt{w^2-1} \, (w+1)^2 
 \Bigg\{\left[1+(1-R_2(w)){w-1 \over 1-r^*} \right]^2 \nn \\ &&
+2\left[{1-2wr^*+r^{*2} \over (1-r^*)^2} \right] \left[1+R_1(w)^2 {w-1 \over w+1} \right] \Bigg\} \,\,\,\,\,\, \label{dGdwexp}
\eea
and $r^*=m_{D^*}/m_B$.
The  function $W_{D^*}(w)$,  defined  in  \cite{Aubert:2008yv}, satisfies the condition $W_{D^*}(1)=(1- \frac{m_\ell^2}{m_B^2(1-r^*)^2})^2 (1+\frac{m_\ell^2}{2 m_B^2(1-r^*)^2})$.
For the three functions in (\ref{dGdwexp}) the parametrization is used \cite{Caprini:1997mu}:
\bea
h_{A_1}(w)&=&h_{A_1}(1) [\, 1-8 {\hat \rho}^2 z+(53 {\hat \rho}^2 -15)z^2  \nn \\ &-& (231{\hat \rho}^2-91)z^3] , \nn \\
R_1(w)&=&R_1(1)-0.12 \, (w-1)+0.05 \, (w-1)^2 , \,\,\,\,\, \label{R1-caprini} \\
R_2(w)&=&R_1(1)+0.11 \, (w-1)-0.06 \, (w-1)^2 ,  \nn
\eea
with 
$
z=\displaystyle{ \sqrt{w+1}-\sqrt{2} \over \sqrt{w+1}+\sqrt{2}} $. 
In  \cite{Aubert:2008yv}  ${\hat \rho}^2$, $R_1(1)$, $R_2(1)$ are fitted separately for $\ell=\mu$ and $\ell=e$. Their values, together with the measurements ${\cal B}(B \to D^* \mu \bar \nu_\mu)=(5.34 \pm 0.06 \pm 0.37)\%$, ${\cal B}(B \to D^*e \bar \nu_e)=(5.50 \pm 0.05 \pm 0.23)\%$,
give
\bea
 h_{A_1}^\mu(1)|V_{cb}|&=&(35.63 \pm 1.96) \times 10^{-3} \label{ha1mu} \\
h_{A_1}^e(1)|V_{cb}|&=&(35.94 \pm 1.65) \times 10^{-3} \label{ha1e}  \,.
\eea
\begin{figure}[h!]
\includegraphics[width = 0.34\textwidth]{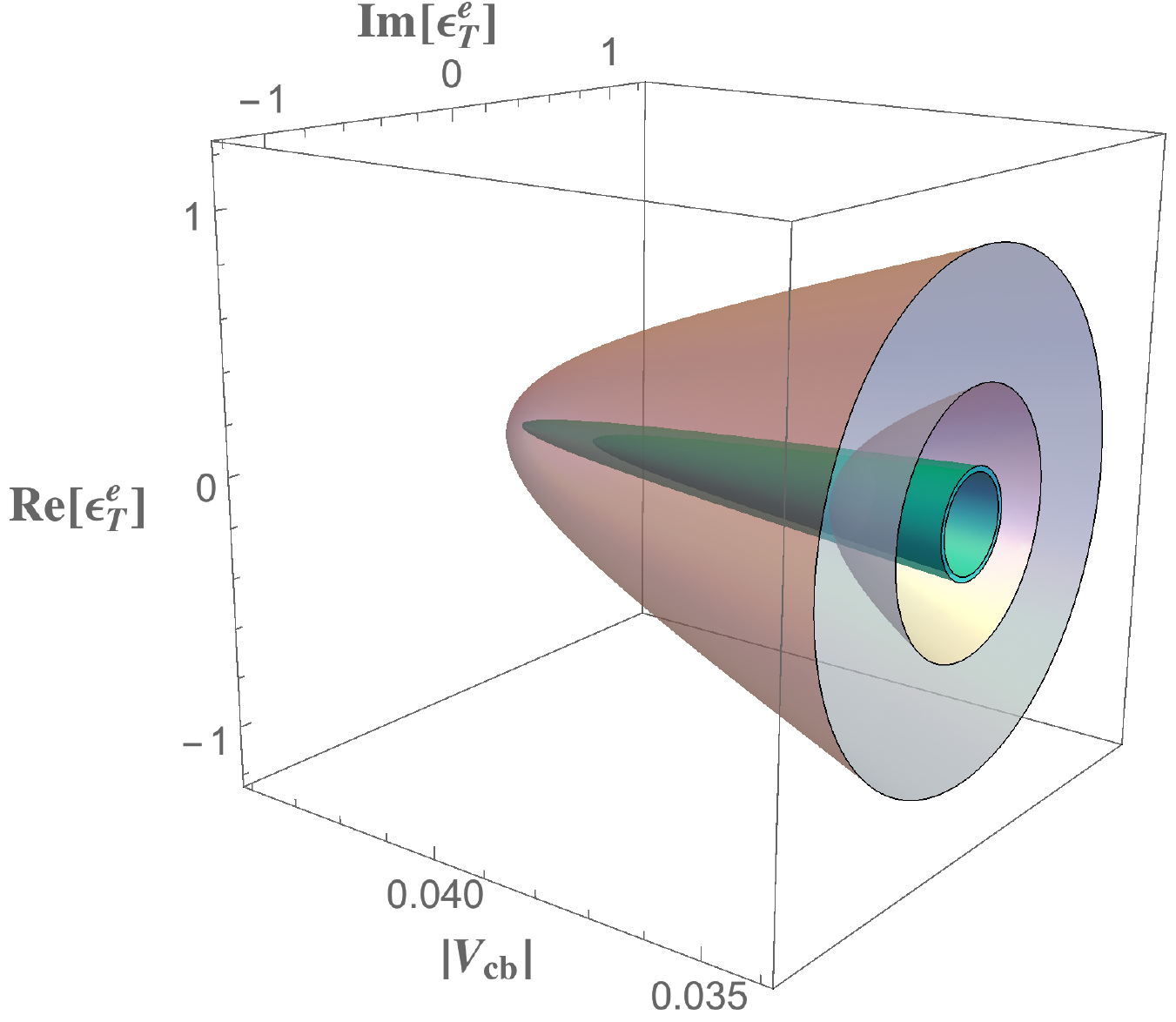}\\
\includegraphics[width = 0.34\textwidth]{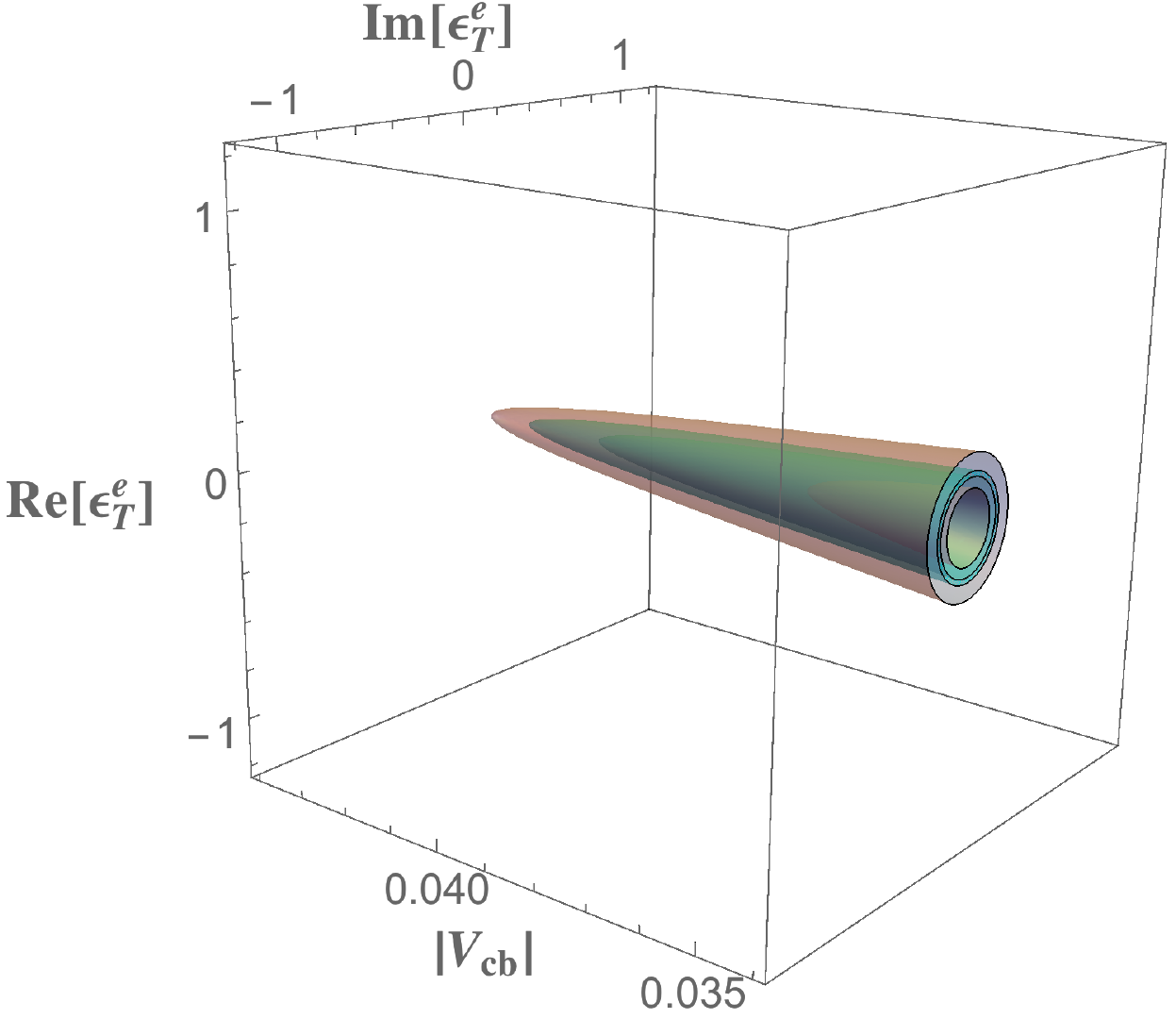}\\
\includegraphics[width = 0.34\textwidth]{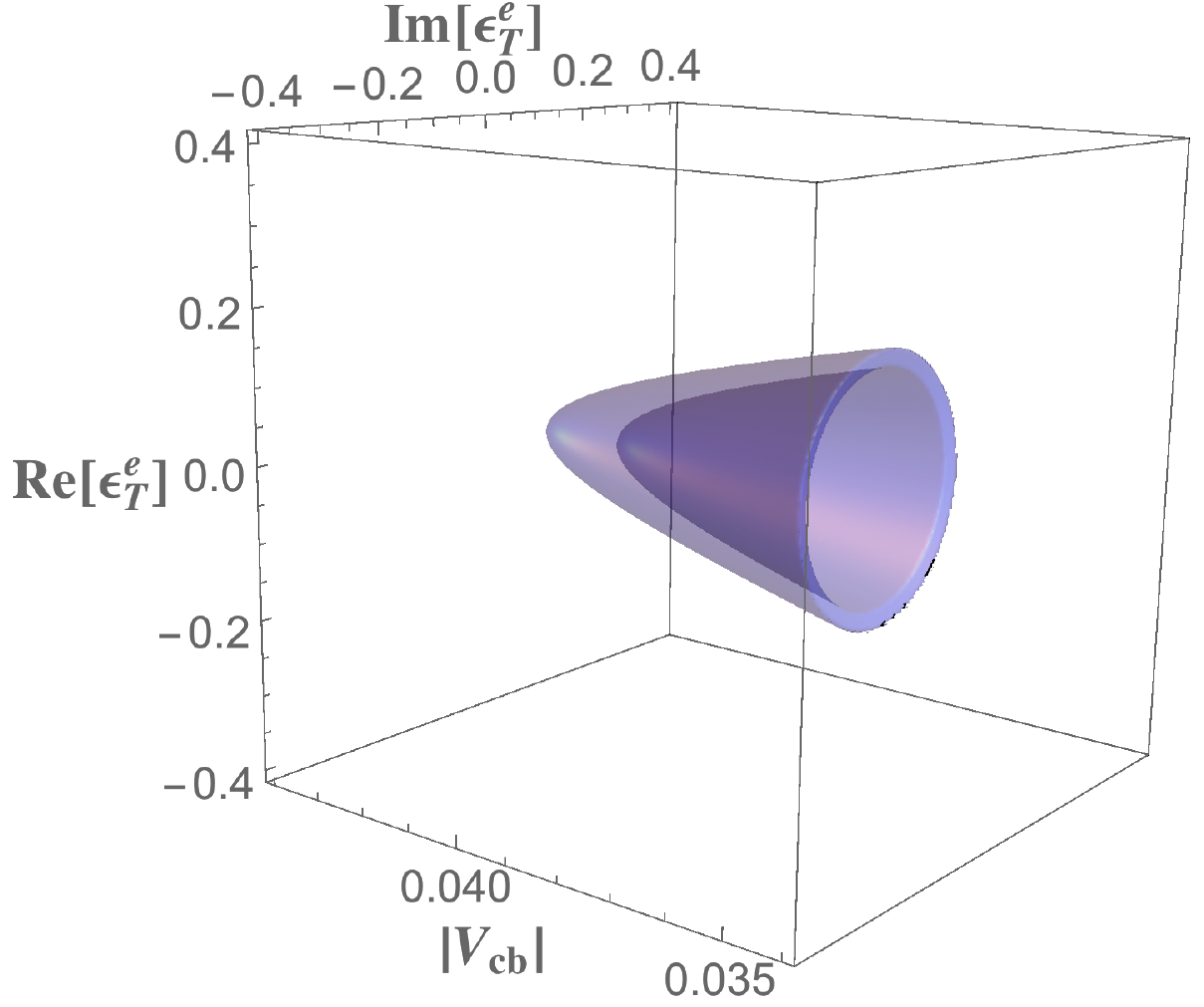}
\caption{Electron modes: allowed regions in the $({\rm Re}(\epsilon_T^e),\,{\rm Im}(\epsilon_T^e),\,\vcb)$ parameter space  from  the inclusive mode (cyan hollow region) together with the  exclusive decay to $D$  (gray hollow region, upper plot), and  from the inclusive mode together with  $D^*$   (gray hollow region, middle plot).   The lower plot 
shows the intersection  of the three regions. }\label{fig:comb-el}
\end{figure}
To bound $({\rm Re}(\epsilon_T^\ell),\,{\rm Im}(\epsilon_T^\ell),\,\vcb)$, we  compare the theoretical expression of  ${d \Gamma}/{dw}$ for $w\to 1$,
\bea
\hskip-0.4cm
&&\frac{d \Gamma^{th}}{dw}(B^- \to D^{*0}  \ell^- {\bar \nu}_\ell)|_{w \to 1} \nn \\
\hskip-0.4cm&&=\frac{G_F^2 \vcb^2}{16 \sqrt{2} \pi^3} \frac{m_{D^*}^2}{m_B}  \sqrt{w-1} \left[1-\frac{m_\ell^2}{(m_B-m_{D^*})^2}\right]^2 
\times \nn \\
\hskip-0.4cm&&\Big\{(m_B+m_{D^*})^2[2(m_B-m_{D^*})^2+m_\ell^2] A_1(1)^2  \label{dGdwth} \\
\hskip-0.4cm&&+|\epsilon_T|^2 4[(m_B-m_{D^*})^2+2 m_\ell^2][m_B {\tilde T}_1(1)+m_{D^*}{\tilde T}_2(1)]^2 \nn \\
\hskip-0.4cm&&
-12 Re(\epsilon_T)(m_B^2-m_{D^*}^2)m_\ell A_1(1)[m_B {\tilde T}_1(1)+m_{D^*}{\tilde T}_2(1)]\Big\}, \nn
\eea
to Eq.~(\ref{dGdwexp}) for $w \to 1$,
using (\ref{ha1mu}) and (\ref{ha1e})  for the muon and the electron mode, respectively. 
In (\ref{dGdwth}),  $\tilde T_i$ are combinations of the tensor form factors
${\tilde T}_0=T_0-T_5,$
${\tilde T}_1=T_1+T_3$ and
${\tilde T}_2=T_2+T_4$.

We are now able to put together the constraints  from the $B$ inclusive  and  exclusive $D, D^*$ decay modes.
\begin{figure}[ht!]
\includegraphics[width = 0.36\textwidth]{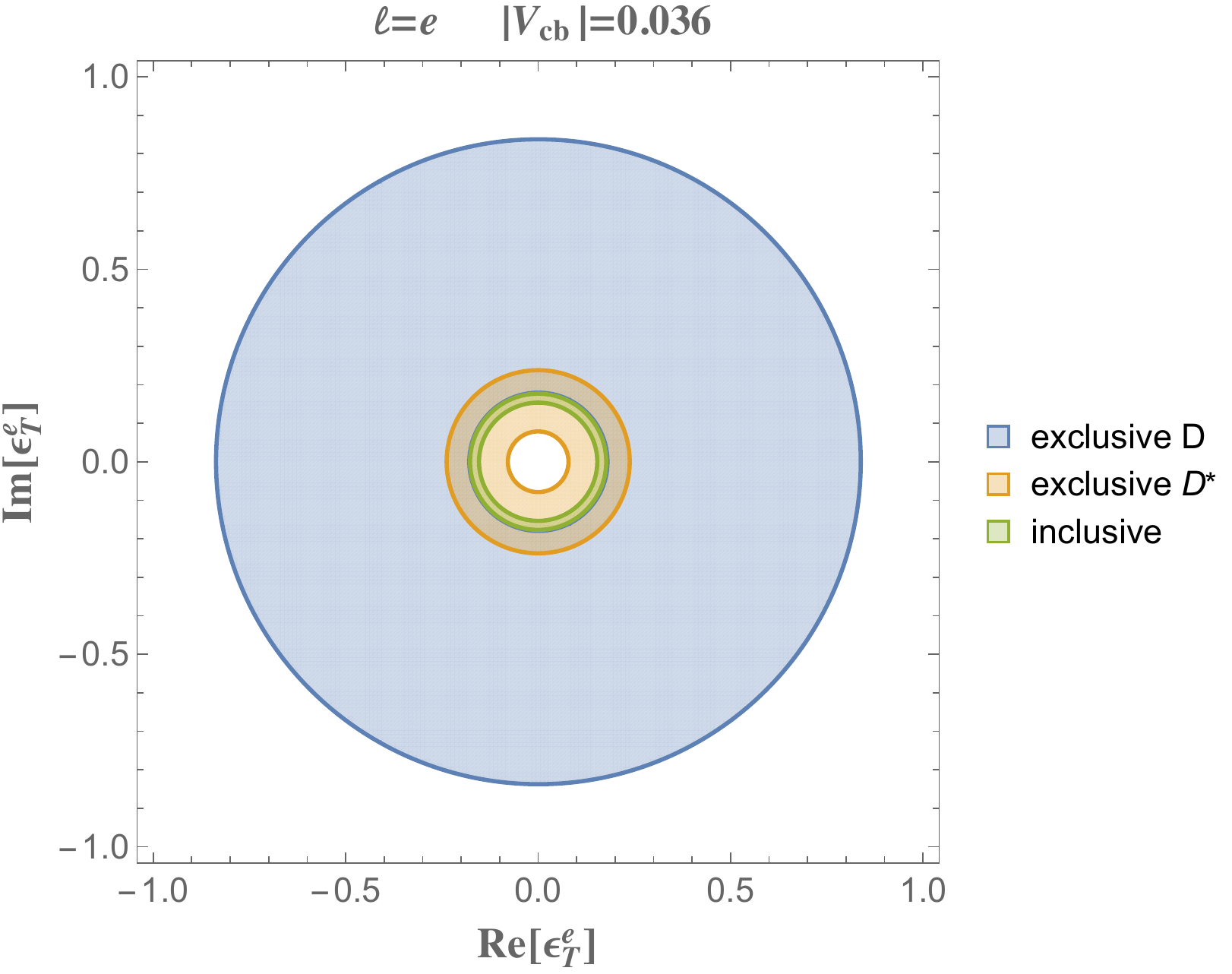}
\caption{Electron channel: projections of the overlap region in Fig.~\ref{fig:comb-el} in the $({\rm Re}(\epsilon_T^e),\,{\rm Im}(\epsilon_T^e))$ plane.  $\vcb$ is set to  the smallest allowed  value. Colour coding as in Fig. \ref{rpnew-mu}.  }\label{rpnew-el}
\end{figure}
For the muon channel,  we show
in the two upper panels of Fig.~\ref{fig:comb-mu}  the parameter regions selected by the exclusive constraints (larger yellow space)   superimposed to the region determined from the inclusive mode  (smaller inner space). The top panel refers to  ${\cal B}(B^- \to D^0 \mu^- \bar \nu_\mu)$,   the middle one to  $\displaystyle{\frac{d \Gamma}{dw}}(B^- \to D^{*0} \mu^- \bar \nu_\mu)$.
In each case there is an overlap between the  regions selected  from the inclusive and the exclusive mode, and a region exists where all  constraints are fulfilled, as shown in the bottom panel. 
The exclusive data slightly reduce the upper bound on $\vcb$ and  produce a lower bound.  This is shown in Fig.~\ref{rpnew-mu}, where we depict the projections in the $({\rm Re}(\epsilon_T^\mu),\,{\rm Im}(\epsilon_T^\mu))$ plane of the parameter space  
corresponding to the extreme values of $\vcb$, i.e., the values for which the parameter regions determined through  the various constraints do not overlap.
After the application of all constraints, the range for $\vcb$ is $\vcb \in [0.0343,\,0.0421]$. The range for $\epsilon_T^\mu$  depends on $\vcb$, and  the largest allowed value is $|\epsilon_T^\mu|\simeq 0.2$.
The symmetry axes of the two regions of parameters bound through the inclusive and the exclusive constraints do not intersect  the $({\rm Re}(\epsilon_T^\mu),\,{\rm Im}(\epsilon_T^\mu))$ plane at the origin and  do not coincide. This is due to the lepton mass effect and to  the interference term in the rates.  The  NP contribution affects in a different way the inclusive  and the two exclusive  $B$ channels.

For  the electron mode  the  analysis is repeated,  with changes in  the results  since the electron mass is tiny.  We show in 
 Fig.~\ref{fig:comb-el} the parameter space selected  from the inclusive and the exclusive $D$ mode (top panel),  and from the inclusive and the  $D^*$ mode (middle panel). The three constraints are fulfilled  for parameters in the region depicted  in the bottom panel. The upper bound on $\vcb$  selected from the inclusive analysis,  $\vcb \le 0.04273$, is not modified by the exclusive constraints, while  a lower bound is  found, as it can be inferred from Fig. \ref{rpnew-el}. The result from the electron modes is  the range $\vcb \in [0.036,\,0.0427]$ and the  bound $|\epsilon_T^e|\le 0.17$.
\begin{figure}[ht!]
\vspace*{-0.3cm}
\hspace*{-0.65cm}
\includegraphics[width = 0.58\textwidth]{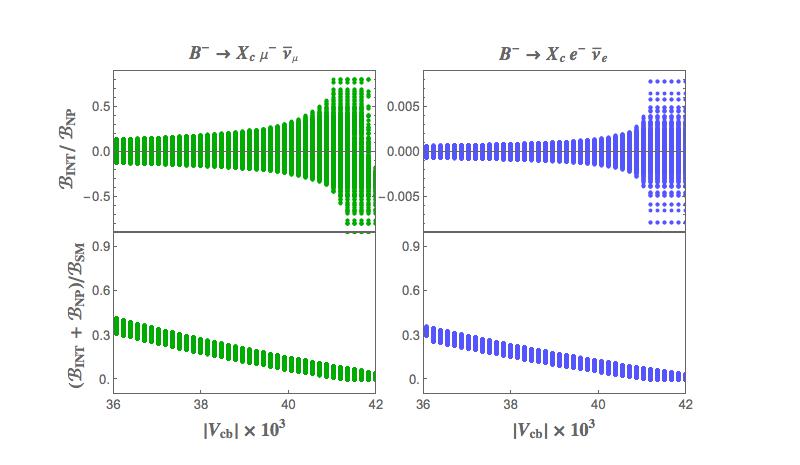}
\vspace*{-0.4cm}
\caption{Relative size of the NP contributions to ${\cal B}(B^- \to X_c \mu^- \bar \nu_\mu)$ (left) and ${\cal B}(B^- \to X_c e^- \bar \nu_e)$ (right).
}\label{grid-incl-mu}
\end{figure}

The conclusion  is that there are sets of three parameters fulfilling all the constraints.  For $\vcb$ this happens in the range  $\vcb \in [0.036,\,0.042]$. Although
 this implies a sizeable uncertainty on $\vcb$, the analysis shows that a  non SM  contribution, as the one considered here,  can reconcile its  inclusive and exclusive  determination.
\begin{figure}[b!]
\vspace*{-0.43cm}
\hspace*{-0.58cm}
\includegraphics[width = 0.58\textwidth]{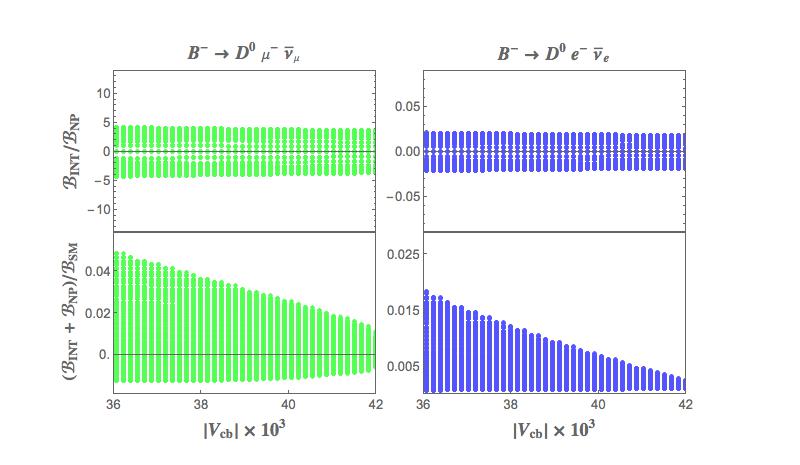}\vspace*{-0.2cm}
\caption{Relative size of the NP contributions to ${\cal B}(B^- \to D^0 \mu^- \bar \nu_\mu)$ (left) and  ${\cal B}(B^- \to D^0 e^- \bar \nu_e)$ (right).
}\label{grid-exclD-mu}
\end{figure}
\\ \indent
To study the role of the NP contribution and of the interference  between  SM and NP,
in the inclusive channel we  separately  integrate the three terms in  (\ref{incl-dgamma}).  Denoting the  resulting quantities, divided by the full decay width,  as ${\cal B}_{SM}$, ${\cal B}_{NP}$, ${\cal B}_{INT}$,   we consider the ratios $\frac{{\cal B}_{INT}}{{\cal B}_{NP}}$ and $\frac{{\cal B}_{INT}+{\cal B}_{NP}}{{\cal B}_{SM}}$, and display   in Fig.~\ref{grid-incl-mu} the results obtained varying $({\rm Re}(\epsilon_T^\ell),\,{\rm Im}(\epsilon_T^\ell),\,\vcb)$ in their allowed region. 
The interference term ${\cal B}_{INT}$ is sizable with respect to ${\cal B}_{NP}$ for muon. Moreover, for both  $\mu$ and $e$, the   NP contribution ${\cal B}_{INT}+{\cal B}_{NP}$ becomes negligible with respect to  SM  for large  $\vcb$, while it is sizable for smaller values of  $\vcb$.  
The analogous quantities for   the exclusive   $ B^- \to D^0 \ell^- \bar \nu_\ell$ modes are in Fig.~\ref{grid-exclD-mu}.
The interference term is again non negligible  for muon.  The impact of the new operator is  larger in the inclusive mode  than in  $D$. The changes due to  NP  are different in the different channels, showing how an extra term in the effective Hamiltonian  could be at the origin of the $\vcb$ anomaly.

 Considering the ranges determined for  the couplings  $\epsilon_T^{e}$ and  $\epsilon_T^{\mu}$,   it is possible that a new physics contribution of the kind considered here  still satisfies  $e-\mu$ universality. Improved measurements and reduced theoretical uncertainties, mainly  in the hadronic form factors, are needed to shed light on this point.

{\it Conclusions.}
We have given an example of a possible mechanism at the origin of  the tension in  $\vcb$ measurements. This mechanism can be tested in other processes, namely  $B_s$   and $\Lambda_b$ semileptonic decays,  for which  enough information is not yet available. A determination of  $\vcb$ from the purely leptonic $B_c \to \ell \bar \nu$ decay is interesting, since in that case the tensor operator in Eq.~(\ref{heff}) does not contribute. As for the experimental investigations,  the  importance of separate measurements of the  muon and  electron  modes cannot be overemphasized.

{\it Acknowledgments.} We thank M. Calvi and M. Rotondo for  discussions. 
This work  has been carried out within the INFN project QFT-HEP.

\bibliography{refs}

\end{document}